\documentclass[twocolumn,showpacs,prl,amsmath,amssymb]{revtex4}
\usepackage{graphicx}
\usepackage{dcolumn}
\usepackage{bm}

\begin{document}

\title{Delocalized fermions in underdoped cuprates}

\author{Mike Sutherland$^{1\star}$, S.Y. Li$^2$, D.G. Hawthorn$^{1\dagger}$, 
R.W. Hill$^{1\ddagger}$, F. Ronning$^{1\S}$, 
M.A. Tanatar$^{1\star\star}$, J. Paglione$^{1\dagger\dagger}$,
H. Zhang$^1$, Louis Taillefer$^{1,2,4}$, J. DeBenedictis$^{3}$, 
Ruixing Liang$^{3,4}$, D.A. Bonn$^{3,4}$ and W.N. Hardy$^{3,4}$}

\affiliation{$^1$Department of Physics, University of Toronto, Toronto,
Ontario, Canada\\
$^2$D\'{e}partement de physique et Regroupement Qu\'{e}b\'{e}cois sur les Mat\'{e}riaux
de Pointe, Universit\'{e} de Sherbrooke, Sherbrooke, Qu\'{e}bec, Canada\\
$^3$Department of Physics and Astronomy, University of British Columbia, Vancouver, British Columbia,
Canada\\
$^4$Canadian Institute for Advanced Research, Toronto, Ontario, Canada}

\date{\today}

\begin{abstract}
 Low-temperature heat transport was used to investigate 
the ground state of high-purity single crystals of the lightly-doped cuprate 
YBa$_{2}$Cu$_{3}$O$_{6.33}$. Samples were measured with doping concentrations
on either side of the superconducting phase boundary.  
We report the observation of delocalized fermionic excitations at zero energy 
in the non-superconducting state, which shows that the ground state of underdoped
cuprates is a thermal metal.  Its low-energy spectrum appears to be similar to that of
the $d$-wave superconductor, i.e. nodal. The insulating ground 
state observed in underdoped La$_{2-x}$Sr$_{x}$CuO$_4$ is attributed to the competing
spin-density-wave order. 
 
 \end{abstract}

\maketitle

Electrons in cuprates adopt a remarkable sequence of ground states as one 
varies the density of charge carriers.  When the electron density in the 
CuO$_2$ planes of a cuprate material is exactly 1.0 per Cu atom in the plane, 
the material is a Mott insulator with static long-range antiferromagnetic 
order, and the electrons are localized on their sites by strong Coulomb 
repulsion. This electronic grid-lock can be relaxed by removing electrons 
from the planes, a charge-transfer process induced by chemical substitution 
away from the planes. This doping process adds $p$ holes per Cu atom in the 
planes, and at high carrier density yields a normal metal with the basic 
signatures of a Fermi liquid.  At intermediate density, it is a 
superconductor with $d$-wave symmetry, but the nature of the underdoped 
phase that lies between the insulator and the superconductor is one of the 
central puzzles of the field.  It is known to be characterised by a 
pseudogap, and is thought to be an exotic state of 
matter \cite{Emery,Chakravarty,Wen,Sachdev}.  As one moves into this 
enigmatic underdoped phase by adding carriers to the Mott insulator a key 
question remains: does the onset of superconductivity coincide with the 
onset of hole mobility? In V$_2$O$_3$ \cite{Limelette} and 2D organic conductors 
\cite{Lefebvre}, pressure studies have answered this question in the affirmative: 
the electron system goes directly from insulator to superconductor, with no 
intermediate phase.

Attempts to address this issue in the cuprates have thus far focused almost 
entirely on the LSCO system \cite{Boebinger}, which is known to be 
intrinsically disordered.  To probe this question in the absence of disorder, 
we turn to the cleanest cuprate available and investigate whether holes are 
mobile or not by measuring their ability to transport heat, as 
$T \rightarrow$ 0. The technique of heat conduction has several 
advantages: 1) it is sensitive only to mobile excitations, 2) it can 
distinguish between fermionic and bosonic excitations, and 3) it probes the 
bulk of a sample, unaffected by possible inhomogeneities at the surface. 
In this Letter, we present a comparative study of heat transport in two 
cuprate materials, YBa$_2$Cu$_3$O$_y$ (YBCO) and La$_{2-x}$Sr$_{x}$CuO$_{4}$ 
(LSCO).  The effect of doping is investigated by 
comparing samples of each material with doping on either side of $p_{SC}$, the critical 
doping for the onset of superconductivity.  
Our main finding is the observation of delocalized fermionic excitations in 
the non-superconducting state of YBCO ($p$ $<$ $p_{SC}$) at $T \rightarrow$ 0.  
This shows that upon doping, a clean cuprate can first go from Mott insulator to thermal metal before it 
turns into a 
superconductor, meaning that holes (or spins) can be mobile without forming a condensate 
of Cooper pairs. In contrast, in the LSCO system, such delocalized low-energy 
excitations are not observed in the non-superconducting state 
\cite{Sun,Hawthorn} showing that excitations are either localized (by the 
stronger disorder) or gapped, possibly by spin-density-wave (SDW) order.

YBCO single crystals of the highest available purity were used, for 
which there is ample evidence of extremely long 
electronic mean free paths \cite{Turner}, estimated to be approximately two 
orders of magnitude longer \cite{Sutherland} than in the best LSCO crystals.
  The oxygen content of an ultra-clean YBCO crystal \cite{Liang}
was set to be near $y=6.33$ so that $p$ would lie 
close to $p_{SC}$. The 
concentration of holes doped into the CuO$_2$ planes depends on the degree 
of oxygen order in the CuO chains, and as a result increases as a function of time spent 
annealing at room temperature \cite{Veal}. The sample was first measured directly 
after growth, before any room-temperature annealing took place.  At this stage it 
was non-superconducting 
(no sign of a transition in resistivity down to $T$= 80 mK) so that $p$ $<$ $p_{SC}$. 
The same sample was re-measured after spending 2 days annealing at room temperature,
after which it was a superconductor with $T_c$= 0.1 K, where we define $T_c$ 
by $\rho$=0.  After
3 weeks of further annealing, $p$ increased such that $T_c$ = 6 K.  For YBCO, 
we use the well-known empirical relation, $T_c/T_c^{max} = 1 - 82.6(p-0.16)^2$ 
\cite{Sutherland}, to define hole concentration. This yields $p$ = 0.051 and 0.054 
for 2 days of annealing and 3 weeks of annealing, respectively.  We 
estimate the non-superconducting state (no annealing) to have $p$ $\sim$ 0.048 by
extrapolating the hole concentration backwards in time on a logarithmic plot 
\cite{Veal}.  For 
LSCO, two samples are used: one non-superconducting ($T_c$ = 0) with $x$=0.05
 ($p$ $<$ $p_{SC}$) and one superconducting ($T_c$  = 5 K) with $x$=0.06 
($p$ $>$ $p_{SC}$), where we simply use $p=x$, the Sr concentration.  

\begin{figure} \centering
\resizebox{\columnwidth}{!}{
\includegraphics{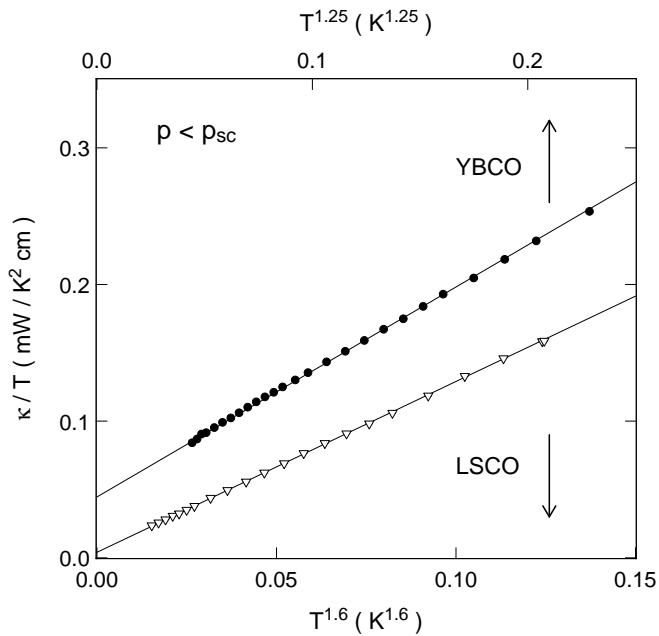}}
\caption{\label{fig:res}Thermal conductivity of underdoped cuprates YBCO and LSCO. 
Both have a hole concentration $p$ close to, but less than $p_{SC}$, the critical 
concentration for the onset of superconductivity. The YBCO sample shows a sizable residual
linear term 
$\kappa_0/T$, indicating the presence of delocalized fermionic carriers of heat. The LSCO sample 
shows a vanishingly small value of $\kappa_0/T$, consistent with an insulating state. }

\end{figure}

We investigate the ground state conductivity by measuring heat transport down  
to 80 mK, which allows for a reliable extrapolation to $T$=0. We begin 
with non-superconducting samples ($p$ $<$ $p_{SC}$). In Figure 1, the thermal conductivity $\kappa$ of 
the non-annealed YBCO sample is compared to that of a previously measured LSCO sample \cite{Hawthorn} with 
$x=0.05$. The data is plotted as $\kappa/T$ vs. $T^{\alpha-1}$ to provide a straightforward way of extrapolating 
to $T=0$, and obtain the residual linear term $\kappa_0/T$. The absence of a residual linear term ($\kappa_0/T$ = 0) 
indicates the absence of fermionic carriers as in an insulator or a fully gapped ($s$-wave) superconductor. A 
finite (non-zero) value can be attributed unambiguously to delocalized fermionic excitations. In either case, the 
slope of the curves is a measure of the phonon conductivity \cite{Sutherland}.  In YBCO, a distinct linear term is 
observed, of magnitude $\kappa_0/T$ = 46 $\pm$ 8 $\mu$W K$^{-2}$ cm$^{-1}$, much larger than that obtained for an 
undoped crystal \cite{Sutherland} ($y$=0.0) where $\kappa_0/T$ = 0 $\pm$ 1 $\mu$W K$^{-2}$ cm$^{-1}$. By contrast, 
the LSCO sample yields a vanishingly small linear term of $\kappa_0/T$ = 3 $\pm$ 1 $\mu$W K$^{-2}$ cm$^{-1}$, just 
as in the undoped ($x=0.0$) material \cite{Hawthorn}, indicating a ground state devoid of delocalized carriers 
for all $p$ $<$ $p_{SC}$ (i.e. $x$ $<$ 0.05).  This observation points to a fundamental difference between 
the two systems: YBCO is a thermal metal, LSCO is a thermal insulator.  Having uncovered delocalized
fermions in a lightly-doped cuprate with no 
long-range superconducting 
order, we explore some of their basic properties: how do they 
compare to the $d$-wave nodal quasiparticles of the superconducting state 
(at $p$ $>$ $p_{SC}$)? How do they respond to a magnetic field?

We begin by comparing the non-superconducting phase below $p_{SC}$ with the coherent superconducting 
state, above $p$ = $p_{SC}$.  The 
dependence of $\kappa_0/T$ on doping is shown in Figure 2, combining present and previous data 
\cite{Sutherland}.  We note that the data is qualitatively similar to the zero-field data reported for $y$ $>$ 6.45 
by Sun $et$ $al.$ \cite{Ando}, although their conclusions differ from ours \cite{footnote}.  In a $d$-wave 
superconductor, nodal quasiparticles give rise to a finite 
$\kappa_0/T$, the magnitude of which is governed entirely by their Dirac energy spectrum.  Indeed, in the universal 
limit, where the residual linear term is independent of impurity concentration, the value of $\kappa_0/T$ only 
depends on the ratio $v_F$/$v_2$, where $v_F$ and $v_2$ are the quasiparticle velocities perpendicular and parallel 
to the Fermi surface, respectively \cite{Durst}:  

\begin{equation}
 \frac{\kappa_0}{T} = \frac{k_B^2}{3 \hbar} \frac{n}{d} \left( 
\frac{v_F}{v_2} + \frac{v_2}{v_F} \right).
 \label{eq:koT}
\end{equation}
  
\noindent Here $n$ is the number of CuO$_2$ planes per unit cell of height $d$. Previous measurements have shown that this 
formalism works remarkably well \cite{Chiao}: in optimally-doped BSCCO, thermal conductivity directly gives 
$v_F$/$v_2$ = 19, while a ratio of $v_F$/$v_2$ = 20 is obtained from independent measurements of $v_F$ and $v_2$ by 
angle-resolved photoemission spectroscopy (ARPES).  

It is straightforward to use such measurements to extract an 
estimate of the superconducting gap maximum, assuming a simple $d$-wave gap of the form $\Delta$ = 
$\Delta_0$~cos~$2~\phi$, so that 2$\Delta_0$ =  $\hbar k_F v_2$.  The value one obtains for the gap in YBCO in this 
manner tracks the value measured by ARPES well into the underdoped regime \cite{Sutherland,Hawthorn2}. This shows that the 
overall decrease in $\kappa_0/T$ with underdoping is caused by a monotonically increasing gap. For the 
highly-underdoped YBCO samples measured here, the linear term of approximately 40 $\mu$W K$^{-2}$ cm$^{-1}$ implies 
a gap maximum of 160 meV, which suggests that the in-plane exchange coupling energy $J$ of the Mott insulator, 
estimated to be 125 meV \cite{Hayden},  sets the magnitude of $\Delta_0$. Note that this type of analysis is only 
valid in the universal limit (i.e. when the scattering rate is small compared to $\Delta_0$), a condition which was 
indeed verified in YBCO at both $y$ = 6.9 and 6.5 \cite{Sutherland}.

\begin{figure} \centering
\resizebox{\columnwidth}{!}{
\includegraphics{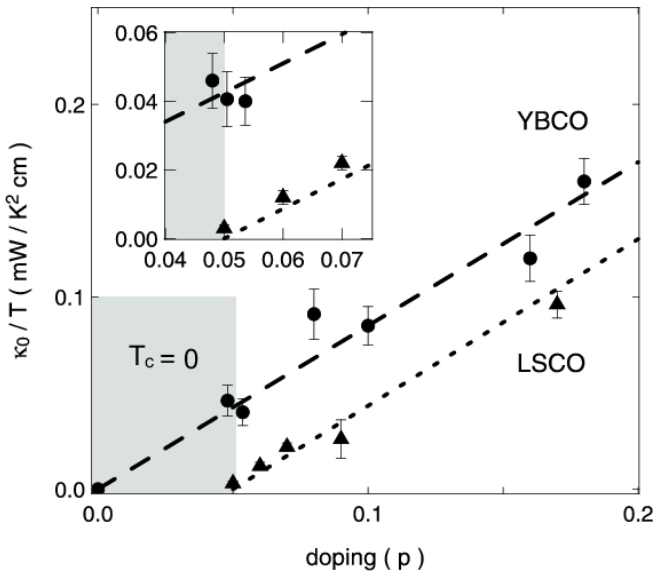}}
\caption{\label{fig:res}   Residual thermal conductivity as a function of carrier density 
$p$ for YBCO (circles) and 
LSCO (triangles). The fermionic heat conductivity decreases with decreasing $p$ in both cases, 
but at $p_{SC}$ it goes 
to zero in LSCO while it remains finite in YBCO.  Dashed lines are guides to the eye. 
Inset: Zoom on region 
close to $p_{SC}$.}

\end{figure}

Two important points emerge from Figure 2. First, the monotonic decrease in $\kappa_0/T$ with underdoping observed in 
YBCO persists smoothly through the critical point at $p$ = $p_{SC}$: there is no detectable change in the 
conductivity of YBCO in going from the $d$-wave superconductor into the non-superconducting phase. Indeed, 
within error bars the residual linear term in the non-annealed state (with $p$ just below $p_{SC}$), $\kappa_0/T$ 
= 46 $\pm$ 8 $\mu$W 
K$^{-2}$ cm$^{-1}$, is identical to the fully-annealed state (with $p > p_{SC}$ and $T_c$ = 6 K), 
$\kappa_0/T$ = 40 $\pm$ 7 $\mu$W K$^{-2}$ 
cm$^{-1}$. Given that the residual linear term is solidly understood as arising from nodal quasiparticles in the 
superconducting state, its seamless evolution into the non-superconducting state below $p_{SC}$ suggests that a nodal 
spectrum is also a characteristic of the thermal metal phase.
 
The second important conclusion one may draw from Figure 2 is that YBCO appears to be qualitatively different from LSCO. 
While in the former the quantum phase transition at $p_{SC}$ has no impact on the conductivity of the electron system, 
in the latter it corresponds to a (thermal) metal-insulator transition. Indeed in LSCO, $\kappa_0/T$ goes to zero 
precisely where superconductivity disappears. The very same situation was observed to occur as a function of applied 
magnetic field, for $p$ $>$ $p_{SC}$: the transition from thermal metal ($d$-wave superconductor) to insulator was found 
to be simultaneous with the suppression of superconductivity, occurring right at the resistive upper critical field 
$H_{c2}$, for a LSCO sample with $x$ = 0.06 \cite{Hawthorn}.

The difference between YBCO and LSCO may lie in the greater amount of disorder found in LSCO, which would cause the 
non-superconducting state of LSCO near $p$ = $p_{SC}$ to be an insulator (thermally and electrically). However, if 
LSCO were merely a disordered version of YBCO, it is hard to see why the metal-to-insulator transition would be pinned 
to the onset of superconductivity (at $p_{SC}$).  The latter fact points instead to another explanation, namely a 
scenario of competing phases where the other phase (e.g. with SDW order) is insulating, for example as a result of 
having a small gap at the nodes \cite{Shen}.  Along these lines, recent neutron scattering studies 
\cite{Lake,Khaykovich} of underdoped LSCO in a magnetic field have revealed a field-induced increase in static SDW 
order. This happens in parallel with the field-induced decrease in conductivity \cite{Hawthorn}. The induced magnetic 
order may well serve to either gap out or localize the fermionic excitations responsible for heat transport as 
$T \rightarrow$  0.

\begin{figure} \centering
\resizebox{\columnwidth}{!}{
\includegraphics{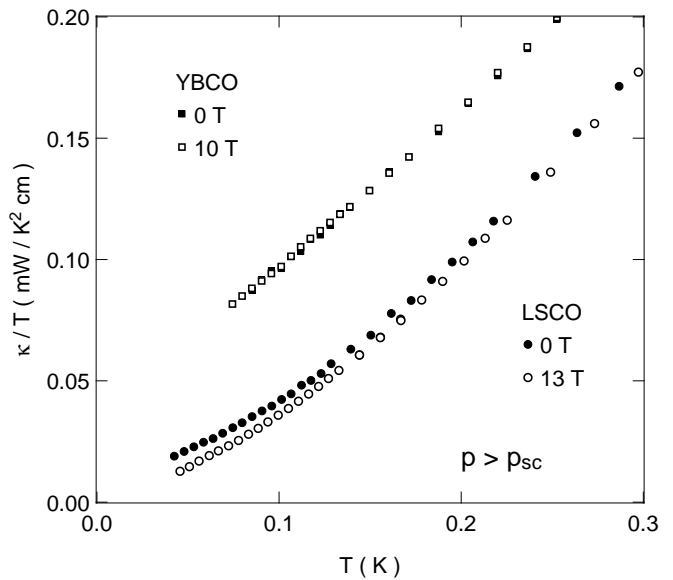}}
\caption{\label{fig:fig3}  Temperature dependence of thermal conductivity for $p$ $>$ $p_{SC}$ in YBCO and 
LSCO, in both the superconducting state (H=0) and the normal state (H$>$H$_{c2}$). The doping $p$ 
is just above $p_{SC}$, $p$=0.051 with $T_c$ = 0.1~K in YBCO and $x$=$p$=0.06 with $T_c$ = 5 K in 
LSCO.}

\end{figure}

Let us now examine the response of YBCO to a magnetic field applied perpendicular to the CuO$_2$ planes. In a $d$-wave 
superconductor, the superfluid flow around each vortex causes a Doppler shift of the quasiparticle energies and thus an 
increase in the zero-energy density of states.  This should lead to an increase in thermal conductivity. In YBCO near 
optimal doping, an increase in $\kappa_0/T$ was observed to be on the order of a factor 2 in 10 T or so \cite{Chiao2}. In
 LSCO, a similar increase is seen at optimal doping, but for $p$ $<$ 0.1, $\kappa_0/T$ was found to decrease, as 
mentioned above and reproduced for $p$=0.06 in Figure 3. This decrease is a signature of the thermal metal-to-insulator 
transition at $H$ = $H_{c2}$. In YBCO, no such decrease is observed for $p$ close to $p_{SC}$  (or anywhere; see \cite{footnote}). In Figure 
3 one can see that the thermal conductivity of the YBCO sample is in fact entirely unaffected by field. 
After 2 days of annealing, the sample has $T_c$ = 0.1 K ($\rho$ = 0) and 10 Tesla is enough to suppress superconductivity 
entirely.  The extrapolated linear term does not change: $\kappa_0/T$ 
= 43 $\pm$ 7 $\mu$W K$^{-2}$ cm$^{-1}$ in 10 T, compared to 40 $\pm$ 7 $\mu$W K$^{-2}$ cm$^{-1}$ in zero 
field.  The same field independence is observed with either $T_c$ = 0 (unannealed) or with $T_c$ = 6 K.  The conclusion is, therefore, that in YBCO near 
$p_{SC}$ the thermal conductivity does not change across the phase boundary, whether one reaches the non-superconducting state 
by decreasing $p$ at fixed $H=0$ or by increasing $H$ at fixed $p$ $>$ $p_{SC}$.

This is reminiscent of previous spectroscopic studies (ARPES \cite{Harris} and tunnelling \cite{Renner}) which found the 
gap in underdoped cuprates to persist largely unchanged as the $temperature$ was increased from below to above $T_c$. 
The observation of this ``pseudogap" above $T_c$ has been interpreted as the persistence of pairing amplitude (gap) once 
long-range superconducting order has been destroyed by thermal fluctuations of the phase \cite{Emery2}. Within such an 
interpretation, the fact that our measurements are done essentially at $T = 0$ would imply a quantum (rather than thermal) 
disordering of the phase with increasing magnetic field or decreasing doping.  What our study shows is that this 
putative phase disordering would leave the system in a ground state with fermionic elementary 
excitations (in this connection see \cite{Franz3}). Beyond this particular interpretation, 
several theoretical models have been proposed for the pseudogap state of underdoped cuprates \cite{Wen,Balents,Franz}.
It remains to be seen which of the proposed states support both a $d$-wave-like gap at high energies and fermionic 
excitations down to zero energy.


In conclusion, we have presented evidence to show that the non-superconducting state of pure cuprates in the underdoped 
regime of the phase diagram is a thermal metal. The associated low-energy fermionic 
excitations have a heat conductivity that 
evolves seamlessly from the superconducting phase, which suggests they have a nodal spectrum akin to that of the 
$d$-wave superconductor. In other words, as holes are doped into the cuprate Mott insulator, a 
thermally metallic ground state is first 
reached before the onset of phase-coherent superconductivity.  We propose this as the generic scenario of clean 
cuprates. Note that it is not realized in the case of LSCO, which instead shows insulating behaviour, most likely caused 
by the presence of a competing SDW order.

\begin{acknowledgments}

We would like to thank S. Fissette, Etienne Boaknin and P. Fournier for their assistance 
in characterizing the samples. This research was supported by NSERC of Canada, a Canada Research Chair (LT) 
and the Canadian Institute for Advanced Research. 

Present addresses: $^{\star}$Cavendish Laboratory, University of Cambridge, Cambridge UK.
$^{\dagger}$Department of Physics and Astronomy, University of British Columbia, Vancouver,
Canada.
$^{\ddagger}$Department of Physics, University of Waterloo, Waterloo, Canada. 
$^{\S}$Los Alamos National Lab, MST-10 Division, Los Alamos, NM, USA. 
$^{\star\star}$Inst. Surf. Chem., N.A.S. Ukraine.
$^{\dagger\dagger}$Department of Physics, University of California, San Diego, La Jolla, CA, USA.


\end{acknowledgments}


\begin{thebibliography}{30}
\bibitem[{foo()}]{footnote}
\bibinfo{note}{The authors claim an insulating ground state in underdoped YBCO
  based on a small suppression of $\kappa$ induced by a magnetic field at
  finite $T$, which is reminiscent of the much larger suppression observed in
  LSCO both at finite $T$ [8], and at $T$ = 0 [9]. We note that a decrease in
  $\kappa$ may occur at finite T even in metallic samples due to
  quasiparticle-vortex scattering, and is not by itself indicative of an
  insulating state (e.g. \cite{Hill}). In our YBCO sample with $y$ = 6.50 ($p$
  = 0.1), we measured a field-induced suppression of $\kappa$ at $T$ = 0.15 K,
  whereas the residual linear term (extrapolated to $T$ = 0) increased with
  field. In this Letter, our claim of a metallic state is based on a more
  direct approach: we simply ask whether there is a sizable residual linear
  term in the non-superconducting state in zero field, accessed by going
  sufficiently low in doping.}
\end{thebibliography}
\end{document}